\documentclass[prl,twocolumn,superscriptaddress,amsmath,amssymb]{revtex4}

\usepackage{hyperref}
\usepackage{amsmath}
\usepackage{amssymb}
\usepackage{graphicx, graphics}
\usepackage{color}
\usepackage{multirow}
\usepackage{dsfont}
\usepackage{mathbbol}

\begin{document}

\title{Full characterization of a carbon nanotube parallel double quantum dot}

\author{Gulibusitan Abulizi}
\email{gulibusitan.abulizi@unibas.ch}
\affiliation{Department of Physics, University of Basel, Klingelbergstrasse 82, CH-4056 Basel, Switzerland}
\author{Andreas Baumgartner}
\affiliation{Department of Physics, University of Basel, Klingelbergstrasse 82, CH-4056 Basel, Switzerland}
\author{C.~Sch\"{o}nenberger}
\affiliation{Department of Physics, University of Basel, Klingelbergstrasse 82, CH-4056 Basel, Switzerland}

\begin{abstract}
We have measured the differential conductance of a parallel carbon nanotube (CNT) double quantum dot (DQD) with strong inter-dot capacitance and inter-dot tunnel coupling. Nominally, the device consists of a single CNT with two contacts. However, we identify two sets of Coulomb blockade (CB) diamonds that do not block transport individually, which suggest that two quantum dots (QDs) are contacted in parallel. We find strong and periodic anti-crossings in the gate and bias dependence, which are only possible if the QDs have similar characteristics. We discuss qualitatively the level spectrum and the involved transport processes in this device and extract the DQD coupling parameters. These results lead us to believe that clean and undoped QDs are formed parallel to the CNT axis, possibly on the outer and inner shells of a multi-wall CNT, or in a double-stranded CNT bundle.
\end{abstract}

\maketitle   

\section{Introduction}
Recently, carbon nanotubes (CNTs) have been used as central elements in a variety of novel electronic devices, owing to their unique electrical and mechanical properties and compatibility with various material types and experimental setups \cite{Laird:2015,Gramich:2015-1,Benyamini:2014,Shulaker:2013,Aurich:2010,Sahoo:2005}. There are many different types of CNTs \cite{Laird:2015}, for example, metallic or semiconducting CNTs. Single-wall CNTs (SWCNTs) are a single sheet of rolled up graphene, while multi-wall CNTs (MWCNTs) consist of several coaxial CNTs of different diameters \cite{Schonenberger:1999,Langer:1996}. In contrast, CNT bundles are a set of separate non-coaxial CNTs in parallel. Long metallic SWCNTs are promising systems, for example, to study one-dimensional Luttinger liquids~\cite{Bockrath:1999}, or novel quasi-particles with non-Abelian statistics \cite{Klinovaja:2013}. SWCNTs of finite length are very reliable in showing size quantization of the energy levels, shell filling effects and Coulomb blockade (CB) in quantum dots (QDs). In comparison to QDs in SWCNTs, MWCNT QDs typically exhibit more complex electronic properties due to more available orbital states, which increase not only the number of conducting channels but also the possibility of intershell interactions \cite{Miyamoto:2001}.

Double QDs (DQDs) are versatile structures that exhibit many physically relevant phenomena \cite{Wiel:2002}. DQDs in series between a source and a drain contact have been investigated regularly \cite{Huttel:2006,Oosterkamp:1998}, also in CNTs \cite{Jung:2013,Graber:2006,Sapmaz:2006}, for example, to investigate spin-blockade \cite{Buitelaar:2008,Liu:2008,Johnson:2005} and quantum bits \cite{Laird:2013}. In parallel DQDs, CB suppresses the electronic transport only if both dots are in blockade. This allows in principle for a more detailed characterization of the individual QDs and the effects of the coupling between the QDs by first order transport processes. However, parallel DQDs are investigated less frequently \cite{Wang:2011} and are more difficult to obtain on CNTs than DQDs in series, because of the close proximity between two CNTs that is required to obtain an appreciable tunnel coupling. Parallel DQDs can in principle form in MWCNTs, where separate QDs might form on different shells, or in non-overlapping parallel CNTs in a bundle, as depicted in Fig.~\ref{Fig:Fabrication}(a) and Fig.~\ref{Fig:Fabrication}(b), respectively. However, if the tunnel coupling is very strong, we expect that the QD states are strongly hybridized and result in the increased degeneracies and shell filling effects typical for MWCNTs~\mbox{\cite{Datta:2011,Moon:2007}}. In contrast, for very small couplings between the concentric CNTs two individual QD characteristics are expected. For intermediate couplings one might expect a hybridization that retains most of the characteristics of the individual QD states, while a pronounced anti-crossing occurs when two charge states become degenerate. Recently, anti-crossings have been observed in a CNT bundle \cite{Gos:2013}, where two or more QDs of very different characteristics have formed. It is difficult in CNT DQDs to gate the QDs individually, so that the DQD characteristics have to be observed in the conductance measured as a function of the bias and a global backgate (BG).

Here, we report differential conductance measurements on a CNT device with two contacts and a global BG at cryogenic temperatures, for which we find two interpenetrating sets of CB diamonds with excited states and strong anti-crossings between specific resonances. These findings are consistent with two strongly coupled parallel QDs in a MWCNT or a tight double-stranded CNT bundle.
 
\section{Device fabrication}
CNTs on substrates often suffer from potential variations on the substrate or residues from the contact fabrication after the CNT growth \cite{Kang:2015,Samm:2014} and cannot be cleaned by current annealing. Here, we employ a stamping method in which CNTs are grown on a separate wafer \cite{Hasler:2015,Viennot:2014} and are later transferred mechanically to the device substrate. The key advantage of CNT stamping techniques is to separate the CNT growth from the fabrication of markers and bonding pads \cite{Gramich:2015-2}. A monolayer chemical vapor deposition (CVD) grown hexagonal boron nitride (hBN) is transferred on top of the CNTs to form a tunnel barrier between the CNTs and the metallic leads. We note that by depositing the hBN layer directly onto the stamped CNTs protects the CNTs from direct exposure to the resist or solvents, which would otherwise contaminate the active structure \cite{Huang:2015,Baumgartner:2014}. Since the hBN layer thickness is small, we believe that the main effect is to reduce the adhesion of residues on the hBN, rather than to increase the distance between the CNTs and the scatterers on the surface.

\begin{figure}[t]
\includegraphics*[width=\linewidth]{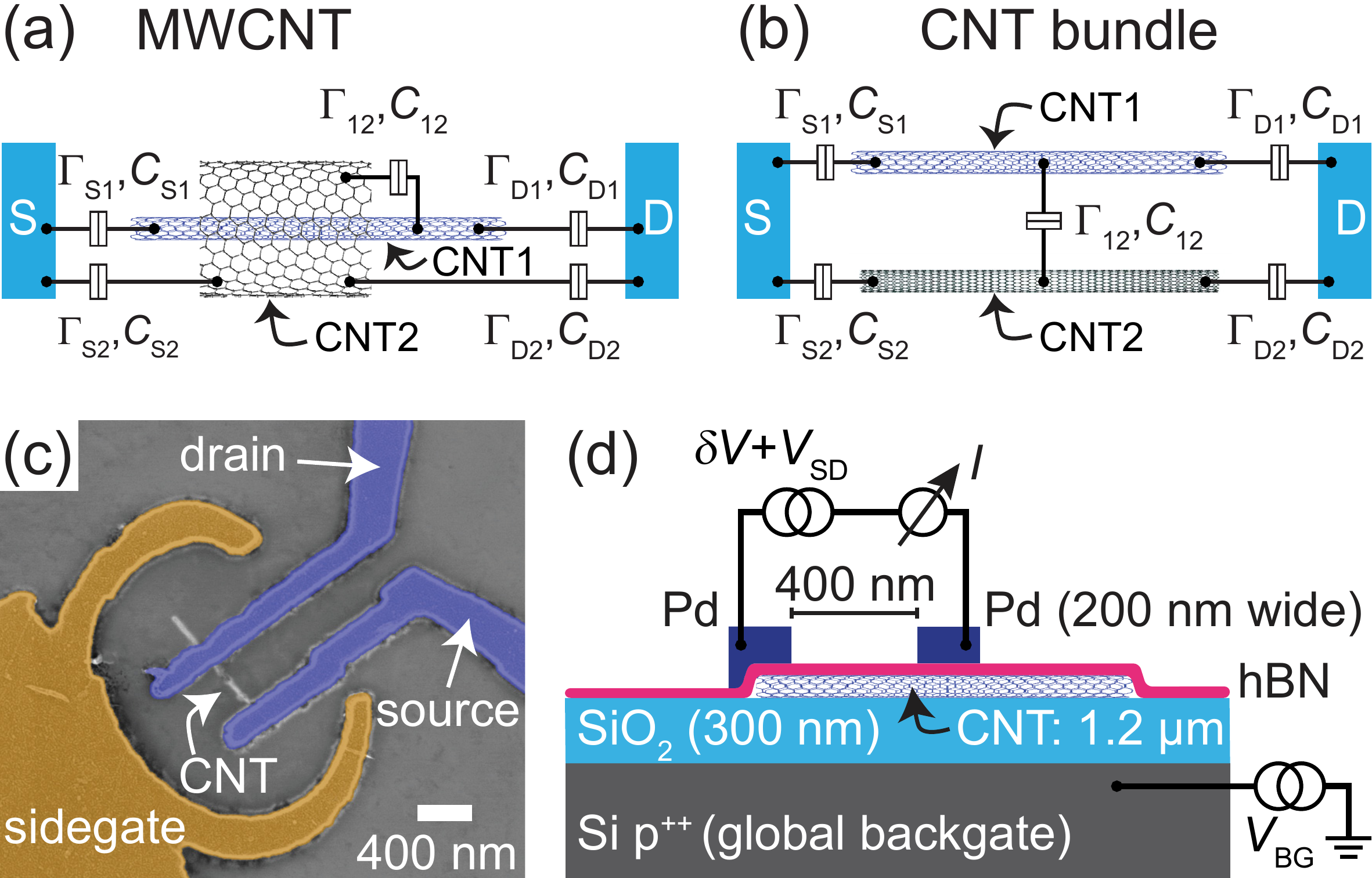}
\caption{Schematic illustration of a parallel-coupled DQDs formed (a) in a MWCNT, or (b) in a CNT bundle. (c) False-color SEM image of a two-terminal CNT device. The CNT is connected to the source and drain electrodes shown in blue and the sidegate shown in orange. (d) Schematic device cross section illustrating the device geometry and electronic setup.}
\label{Fig:Fabrication}
\end{figure}

The device fabrication starts with the manufacturing of the CNT stamps. A silicon (Si) substrate capped by a thermal oxide (SiO$_2$) layer is patterned into an array of square mesas using electron-beam lithography. Each mesa is $50\,\mathrm{\mu m}$ long and wide and $4\,\mathrm{\mu m}$ high, with a spacing of $50\,\mathrm{\mu m}$ between neighboring squares. After spin coating of Fe/Mo catalyst particles onto the mesas, we grow CNTs at $950\,^{\circ}$C for 10 minutes in a CVD process with methane as carbon precursor gas. The target substrate is a piece of a heavily p-doped Si wafer with $300\,$nm thick SiO$_2$ on top, which acts as a global BG. This substrate is patterned with $5\,$nm/$45\,$nm Ti/Au markers and bonding pads. We then transfer the CNTs from the mesa substrate onto the target substrate using a mask aligner, by which the mesa and the target substrates are roughly aligned using the markers and pressed together. About $6-10$ CNTs are transferred to a $200\times200\,\mathrm{\mu m}^2$ area on average. We locate the CNTs using a scanning electron microscope (SEM) \cite{Baumgartner:2014}. Immediately after this step a monolayer CVD hBN (from Graphene Supermarket) is transferred by a wet-etch process from its growth substrate to the target substrate \cite{Li:2009} with the CNTs below. Thermal annealing at $200\,^{\circ}$C for 2 hours removes the poly(methyl methacrylate) (PMMA) resist residues on top of the transferred hBN layer \cite{Longchamp:2013}. Suitable CNTs are then contacted by $10\,$nm/$50\,$nm Cr/Pd source and drain contacts using electron-beam lithography. 

An SEM image of the resulting device is shown in Fig.~\ref{Fig:Fabrication}(c), and a schematic cross section with details of the device geometry and the electrical measurement setup is shown in Fig.~\ref{Fig:Fabrication}(d). A $1.2\,\mathrm{\mu m}$ long CNT is contacted by $200\,$nm wide electrodes, separated by $400\,$nm. One contact (source) covers the end of the CNT, while the other (drain) does not. In this device a circular sidegate (SG) is fabricated in the same step for additional tunability. The sidegate voltage is kept constant for this work and will not be discussed further.

\section{Electrical characterization}

At room temperature the device has a resistance of $5\,\mathrm{M\Omega}$ for negative gate voltages ($V_{\rm BG}\approx {-2\,}$V). Low-temperature transport properties are characterized in a $^3$He refrigerator at a base temperature of $\sim{245}\,$mK. We apply a dc and an ac bias, $V_{\rm SD}+\delta V$, to one contact (source) and measure the differential conductance $G=\delta I/\delta V$ of the device using standard lock-in techniques with an ac voltage of $\delta V = 40\,\mathrm{\mu V}$ at a frequency of $328\,$Hz, as illustrated in Fig.~\ref{Fig:Fabrication}(d).

\begin{figure}[t]
\includegraphics*[width=\linewidth]{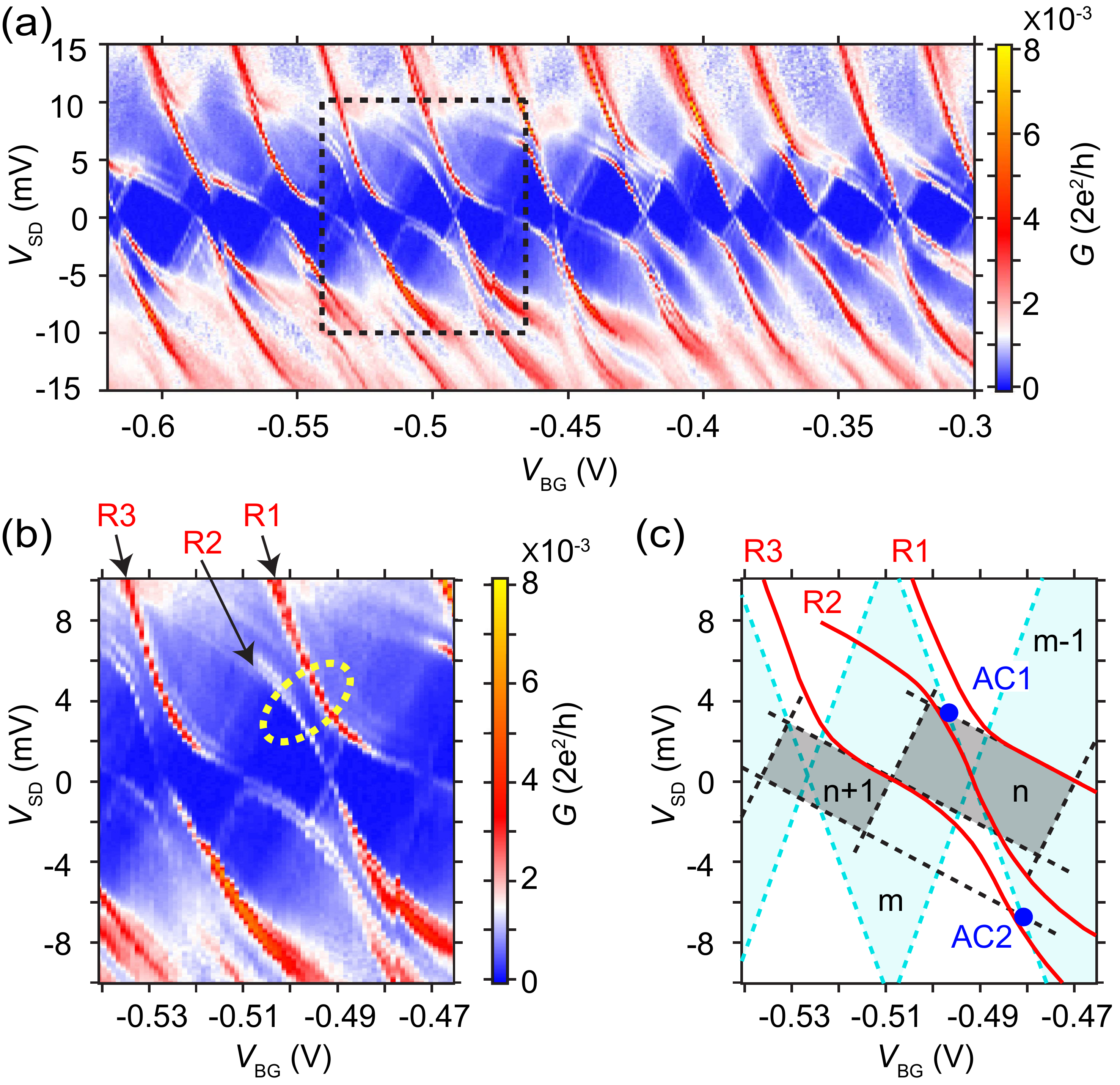}
\caption{(a) Colorscale plot of the differential conductance $G$ as a function of $V_\mathrm{BG}$ and $V_\mathrm{SD}$ for a fixed SG voltage $V_\mathrm{SG}=-2\,$V and at $T=245\,$mK. (b) Magnification of the region indicated in Fig.~\ref{Fig:Electrical Characterization}(a). Three resonances are labeled as R1, R2, and R3. The yellow dashed ellipse highlights an avoided crossing. (c)~Schematic charge stability diagram of the parallel DQDs extracted from Fig.~\ref{Fig:Electrical Characterization}(b). The three solid red lines correspond to the transport resonances (R1, R2, and R3) marked in Fig.~\ref{Fig:Electrical Characterization}(b) and the dashed lines are the extrapolated lines that separate the charge states of the individual QDs. AC1 and AC2 point out two avoided crossings, while $n$ and $m$ are the number of holes in the respective charge states.}
\label{Fig:Electrical Characterization}
\end{figure}

Figure~\ref{Fig:Electrical Characterization}(a) shows a colorscale plot of $G$ as a function of $V_\mathrm{BG}$ and $V_\mathrm{SD}$. We find a periodic pattern of strongly distorted CB diamonds, suggesting the formation of QDs in the CNT. While the weak CB diamond boundaries with positive slopes are straight, the ones with negative slopes consist of a series of avoided crossings. These lines have a larger amplitude, especially at larger bias, can be rather wide and can even occur in pairs.

To characterize the QDs formed in the device, we focus on the region pointed out by the dashed rectangle, with the corresponding data replotted in Fig.~\ref{Fig:Electrical Characterization}(b), while in Fig.~\ref{Fig:Electrical Characterization}(c) the positions of the resonances R1, R2 and R3 of Fig.~\ref{Fig:Electrical Characterization}(b) are plotted as solid red lines. First we identify individual CB diamonds and ignore the avoided crossings and other effects discussed below. For this we extrapolate the resonance position around zero bias, which results in the dashed black and blue diamonds. This two-QD pattern is shown exemplarily in Fig.~\ref{Fig:Electrical Characterization}(c), but also describes roughly the extended data set of Fig.~\ref{Fig:Electrical Characterization}(a). We therefore conclude that two QDs are formed in the system.

From these extrapolated CB diamonds we estimate the charging energies of the two individual QDs as $E_\mathrm{C1}\approx{10.4\,}$meV and $E_\mathrm{C2}\approx{3.0\,}$meV, which correspond to the total capacitances $C_\mathrm{tot1}\approx{15\,}$aF and $C_\mathrm{tot2}\approx{53\,}$aF, respectively. We label the QD with the larger charging energy as QD1 and the other as QD2. In the constant interaction model \cite{Kouwenhoven:2001} the positive and negative slopes of an individual CB diamond are given by $+\,\frac{C_\mathrm{BG}}{C_\mathrm{tot}-C_\mathrm{S}}$ and $-\,\frac{C_\mathrm{BG}}{C_\mathrm{S}}$, with $C_\mathrm{tot}=C_\mathrm{BG}+C_\mathrm{S}+C_\mathrm{D}$. From these expressions we obtain the capacitances listed in Table~\ref{table:coupling parameters} for the individual QDs. Here, we neglect the inter-dot capacitance, which might explain the small discrepancies in the sums from the measured $C_{\mathrm{tot}}$. We find very similar values for the capacitive coupling of the drain to both QDs, for the BG to both QDs, and for the source to QD1. However, $C_{\rm S}$ of QD2 is about $8$ times larger, possibly related to the fact that this is the contact that overlaps the end of the CNT. We note that both BG capacitances are virtually identical. In addition, we can use the full width at half maximum of the zero-bias resonances as upper limits for the tunnel coupling strengths, yielding $\mathrm{\Gamma}_\mathrm{1}\leq{460\,\mathrm{\mu eV}}$ and $\mathrm{\Gamma}_\mathrm{2}\leq{305\,\mathrm{\mu eV}}$, respectively.


\begin{table}[b]
   \caption{Extracted parameters for QD1 and QD2.}
	 \label{table:coupling parameters}
   \begin{tabular}{@{}llll@{}}
     \hline
     parameters &  QD1 (blue lines)  && QD2 (black lines) \\
     \hline
     $C_\mathrm{tot}$ & $15.3\,$aF & & $53.3\,$aF   \\
     $C_\mathrm{BG}$ & $5.0\,$aF & & $5.3\,$aF   \\
     $C_\mathrm{S}$  & $8.1\,$aF & & $41.0\,$aF  \\
     $C_\mathrm{D}$  & $2.6\,$aF & & $6.0\,$aF  \\
     $\mathrm{\Gamma}$  & $460\,\mathrm{\mu eV}$ & & $305\,\mathrm{\mu eV}$  \\
     $C_{12}$				& & $\sim 5\,$aF & \\
     $\mathrm{\Gamma_{12}}$				& & $\geq500\,\mathrm{\mu eV}$ & \\
     \hline
   \end{tabular}
 \end{table} 

Figure~\ref{Fig:Electrical Characterization}(a) also shows excited state resonances, which run in parallel to the CB diamond boundaries. These lines are most pronounced for the resonances with negative slopes, which suggest fairly asymmetric tunnel barriers~\cite{Kouwenhoven:2001}. We extract the mean energy difference between these resonances as $\delta E\approx 0.9\,$meV. If we assume the confinement length $L$ to be the $400$ nm spacing between the source and drain electrodes, we expect a mean level spacing $\delta{E}=hv_\mathrm{F}/2L\approx{4}\,$meV for an ideal and undoped metallic CNT, with $h$ the Plank constant and $v_\mathrm{F}=8\times{10^5}\,$m/s the Fermi velocity \cite{Lemay:2001}. This expected value is a factor of four larger than the energy difference between the excited states in Fig. \ref{Fig:Electrical Characterization}(a), suggesting both QDs are considerably larger than the contact spacing. Though we might overestimate the level spacing in case of a semiconducting CNT because of the flat electronic band structure close to the band gap \cite{Cao:2005-2}, we speculate that by introducing a monolayer hBN tunnel barrier between the selected CNT and the metal contacts, it is possible that a larger QD forms on the significantly longer CNT, because the contacts are weakly coupled and do not necessarily result in electron confinement. As a result, the QD wave function can extend beyond the spacing between the source and drain contacts for weakly coupled tunnel contacts (hBN layer). 

\begin{figure}[t]
\centering
\includegraphics*[width=\linewidth]{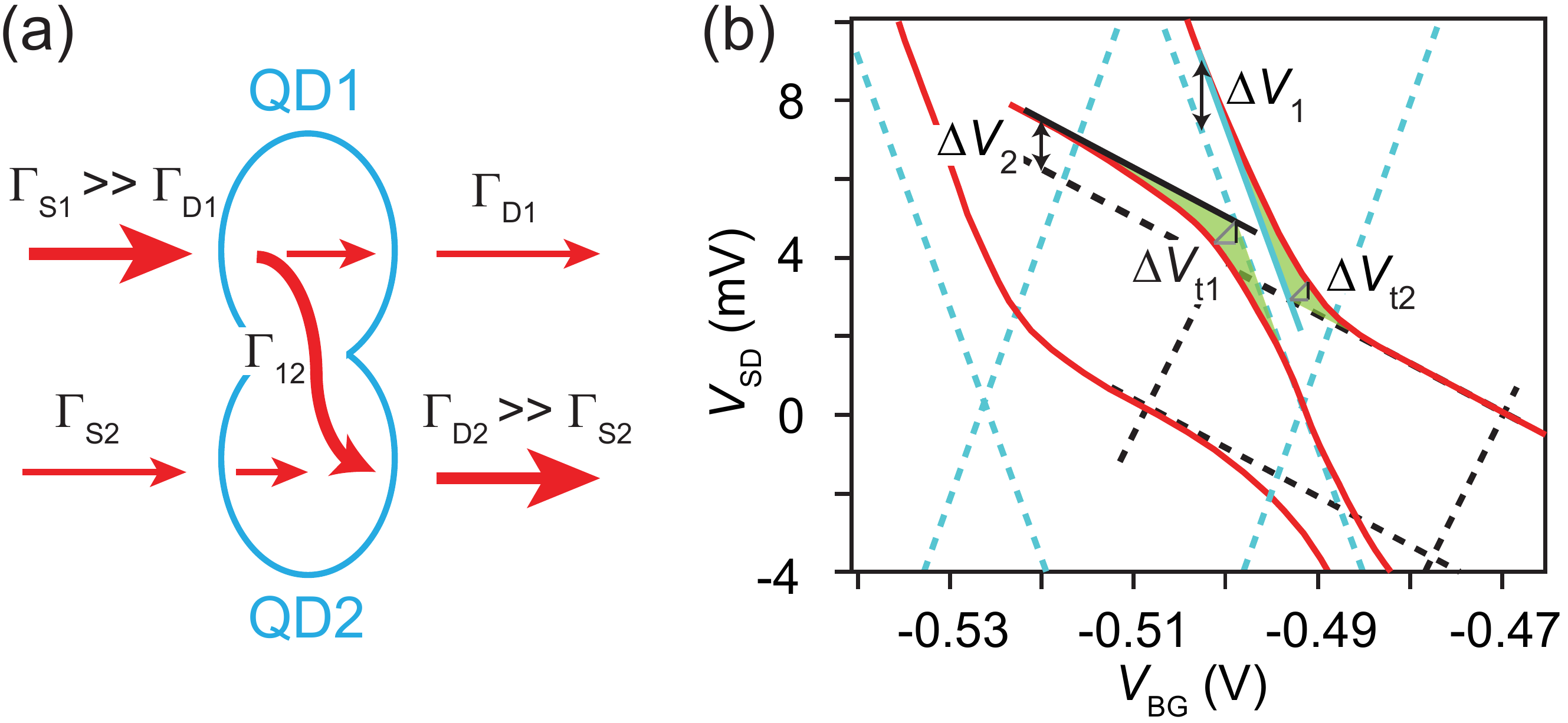}
\caption{(a) Schematic sequential tunneling through a strongly coupled DQD. (b) Replot of the charge stability diagram in Fig.~\ref{Fig:Electrical Characterization}(c) with the focus on the avoided crossing AC1 discussed in the text.}
\label{Fig:charge states}
\end{figure}

We now turn to the discussion of the avoided crossings shown for example in Fig.~\ref{Fig:Electrical Characterization}(b), where the avoided crossing AC1 is highlighted by a yellow dashed ellipse.  An avoided crossing is observed at the intersection points between the CB diamond boundaries of QD1 and QD2 with negative slopes. This can be understood easily by considering that at these points the chemical potentials ("resonances") of both QDs would both be aligned with the electrochemical potential of the drain ($\mathrm{\mu_D=0}$), which means that both QD potentials are identical and electrons (or holes) can be exchanged not only with the leads, but also between the QDs. This results in a hybridization of the QD wave functions and an avoided crossing in their spectrum. The increased resonance amplitudes can be understood qualitatively in the sequential tunneling picture by considering the case $\mathrm{\Gamma_{12}}\gg\mathrm{\Gamma}_{j}$, where $j$ stands for all the contacts. As illustrated in Fig.~\ref{Fig:charge states}(a) the DQD then acts like a single QD with $4$ leads. In addition to the paths through the individual QDs, electrons (or holes) can also tunnel into one QD and out of the other, which can result in more than the sum of the currents through the individual QDs. The total tunneling rate reads $\mathrm{\Gamma}=(\mathrm{\Gamma_{S1}}+\mathrm{\Gamma_{S2}})(\mathrm{\Gamma_{D1}}+\mathrm{\Gamma_{D2}})/\mathrm{\Gamma_{\Sigma}}=(\mathrm{\Gamma_{S1}}\mathrm{\Gamma_{D1}}+\mathrm{\Gamma_{S2}}\mathrm{\Gamma_{D2}}+\mathrm{\Gamma_{S2}}\mathrm{\Gamma_{D1}}+\mathrm{\Gamma_{S1}}\mathrm{\Gamma_{D2}})/\mathrm{\Gamma_{\Sigma}}$ with $\mathrm{\Gamma_{\Sigma}}\approx \mathrm{\Gamma_{S1}}+ \mathrm{\Gamma_{S2}}+\mathrm{\Gamma_{D1}}+\mathrm{\Gamma_{D2}}$. The first two terms are essentially the individual QD transmissions, which are dominated by the last two terms for the situation of very asymmetric couplings shown in Fig.~\ref{Fig:charge states}(a). We would in principle expect a similar effect for the CB resonances with positive slopes (the dot potentials aligned with the source Fermi energy). However, since the two positive slopes are very similar, no such crossing is observed on this device.

To characterize the avoided crossings, we replot the positions of the three resonance curves R1, R2 and R3 in Fig.~\ref{Fig:charge states}(b) and focus on the avoided crossing AC1. We now draw the asymptotes to the resonances away from AC1. Two lines are the CB diamond edges found above, but the other two are offset in bias by $\mathrm{\Delta} V_1$ and $\mathrm{\Delta} V_2$, respectively. These offsets are in analogy with the zero-bias gate maps in more standard, separately gated DQDs in series \cite{Wiel:2002}. The offsets are due to one QD capacitively sensing the charge state of the other, while the bending of the resonances, indicated as green shadings in Fig.~\ref{Fig:charge states}(b), stems from the inter-dot tunnel coupling.

We first extract the inter-dot capacitance $C_{12}$ from the resonance offsets: the addition of one electron to QD~$i$ results in a change of the electrical potential in QD $j$, $\mathrm{\Delta}\Phi_j$ (and vice versa), due to the capacitive coupling. For $C_{12}\ll C_{\mathrm{tot}1},C_{\mathrm{tot2}}$ one finds $\mathrm{\Delta}\Phi_j=\frac{e^2}{C_\mathrm{tot1}C_\mathrm{tot2}}C_{12}$. This shift in the potential has to be compensated by a change $\mathrm{\Delta} V^{(j)}_{\rm SD}$ in the bias measured between the two asymptotes corresponding to QD $j$. For the drain resonance one obtains $0=\mathrm{\Delta}\Phi_j + e\alpha_{\mathrm{S}j}\mathrm{\Delta} V_{\rm SD}$ with the source lever arm $\alpha_{\mathrm{S}j}=\frac{C_{\mathrm{S}j}}{C_{\mathrm{tot}j}}$. From this, one directly obtains:
\begin{equation}
C_{12}=-\frac{C_{\mathrm{tot}i}C_{\mathrm{S}j}}{e}\mathrm{\Delta} V^{(j)}_{\rm SD}
\end{equation}
Similarly, from the the resonance condition at the source contact, $e\mathrm{\Delta} V^{(j)}_{\rm SD}=\mathrm{\Delta}\Phi_j + e\alpha_{\mathrm{S}j}\mathrm{\Delta} V^{(j)}_{\rm SD}$, one then obtains $C_{12}=\frac{1}{e}C_{\mathrm{tot}i} (C_{\mathrm{tot}j} - C_{\mathrm{S}j})\mathrm{\Delta} V^{(j)}_{\rm SD}$.
Inserting the experimental values for the offsets $\mathrm{\Delta} V^{(j)}_{\rm SD}$ and the capacitances in Table~\ref{table:coupling parameters}, we find consistently for both QD resonance lines at AC1 the inter-dot capacitance $C_{12}\approx5\,$aF. Interestingly, this value varies between $5\,$aF and $9\,$aF for four neighboring avoided crossings, which might be either due to other crossings nearby (here, for example AC2), or a deeper reason, possibly due to the QD quantum capacitance that might change with the charge and orbital states, and with the bias. The extracted value is comparable to the gate and contact capacitances, so that this value has to be taken as an approximation.

We estimate the inter-dot tunnel coupling strength by considering only the bias component of the bending, $\mathrm{\Delta} V_{\rm t}$, as illustrated in Fig.~\ref{Fig:charge states}(b). This results in a lower limit for the tunnel coupling, $\mathrm{\Gamma_{12}}\geq500\,\mathrm{\mu eV}$. We note that we find a rather large variation ($\sim 20$\%) between the extracted values for different avoided crossings, which probably originates from errors in the asymptotic lines. We point out that this value is of similar strength as the total tunnel coupling to the leads.

One might expect that with the inter-dot coupling parameters one should be able to distinguish whether the DQD is formed on two shells of a MWCNT or on two separate CNTs in a bundle, see Fig.~\ref{Fig:Fabrication}(a) and Fig.~\ref{Fig:Fabrication}(b), respectively. The expressions for the capacitances of two parallel or coaxial cylinders at a distance compatible with a large inter-dot tunnel coupling (few nanometers) both suggest unphysically small CNT diameters. The reason for this is that the source and drain contacts reduce the inter-dot capacitance due to screening, which can only be accounted for numerically \cite{Baumgartner:2014}.  However, two parallel CNTs in a bundle would naturally account for the identical gate capacitances of the two QDs.

\section{Conclusions}
We report low-temperature differential conductance measurements on parallel DQDs, formed on two shells of a MWCNT or on two individual CNTs of a bundle. We investigate avoided crossings that result from the tunnel and capacitive couplings between the electronic charge states of different QDs. Our results enrich the fundamental understanding of quantum transport through coupled QDs formed in a parallel configuration. We demonstrate that in the sense of the simplest DQD model (large level spacing and constant interaction), transport spectroscopy can be used as a sensitive tool to fully characterize the interactions between parallel-coupled QDs also in a two-terminal CNT device with only a single global gate. 

\section{Acknowledgements}
We thank T.~Hasler and K.~Thodkar for their help with the CNT stamps and the  hBN transfer process, respectively. This work was financially supported by the Swiss National Science Foundation (SNF), the Swiss Nanoscience Institute (SNI), the Swiss NCCR QSIT, and the European ERC project QUEST.

\bibliography{CNT_DQD_2016}

\begin{thebibliography}{37}
\expandafter\ifx\csname natexlab\endcsname\relax\def\natexlab#1{#1}\fi
\expandafter\ifx\csname bibnamefont\endcsname\relax
  \def\bibnamefont#1{#1}\fi
\expandafter\ifx\csname bibfnamefont\endcsname\relax
  \def\bibfnamefont#1{#1}\fi
\expandafter\ifx\csname citenamefont\endcsname\relax
  \def\citenamefont#1{#1}\fi
\expandafter\ifx\csname url\endcsname\relax
  \def\url#1{\texttt{#1}}\fi
\expandafter\ifx\csname urlprefix\endcsname\relax\def\urlprefix{URL }\fi
\providecommand{\bibinfo}[2]{#2}
\providecommand{\eprint}[2][]{\url{#2}}

\bibitem[{\citenamefont{Laird et~al.}(2015)\citenamefont{Laird, Kuemmeth,
  Steele, Grove-Rasmussen, Nyg{\aa}rd, Flensberg, and
  Kouwenhoven}}]{Laird:2015}
\bibinfo{author}{\bibfnamefont{E.~A.} \bibnamefont{Laird}},
  \bibinfo{author}{\bibfnamefont{F.}~\bibnamefont{Kuemmeth}},
  \bibinfo{author}{\bibfnamefont{G.~A.} \bibnamefont{Steele}},
  \bibinfo{author}{\bibfnamefont{K.}~\bibnamefont{Grove-Rasmussen}},
  \bibinfo{author}{\bibfnamefont{J.}~\bibnamefont{Nyg{\aa}rd}},
  \bibinfo{author}{\bibfnamefont{K.}~\bibnamefont{Flensberg}},
  \bibnamefont{and} \bibinfo{author}{\bibfnamefont{L.~P.}
  \bibnamefont{Kouwenhoven}}, \bibinfo{journal}{Reviews of Modern Physics}
  \textbf{\bibinfo{volume}{87}}, \bibinfo{pages}{703} (\bibinfo{year}{2015}).

\bibitem[{\citenamefont{Gramich
  et~al.}(2015{\natexlab{a}})\citenamefont{Gramich, Baumgartner, and
  Sch\"onenberger}}]{Gramich:2015-1}
\bibinfo{author}{\bibfnamefont{J.}~\bibnamefont{Gramich}},
  \bibinfo{author}{\bibfnamefont{A.}~\bibnamefont{Baumgartner}},
  \bibnamefont{and}
  \bibinfo{author}{\bibfnamefont{C.}~\bibnamefont{Sch\"onenberger}},
  \bibinfo{journal}{Physical Review Letters} \textbf{\bibinfo{volume}{115}},
  \bibinfo{pages}{216801} (\bibinfo{year}{2015}{\natexlab{a}}).

\bibitem[{\citenamefont{Benyamini et~al.}(2014)\citenamefont{Benyamini, Hamo,
  Kusminskiy, von Oppen, and Ilani}}]{Benyamini:2014}
\bibinfo{author}{\bibfnamefont{A.}~\bibnamefont{Benyamini}},
  \bibinfo{author}{\bibfnamefont{A.}~\bibnamefont{Hamo}},
  \bibinfo{author}{\bibfnamefont{S.~V.} \bibnamefont{Kusminskiy}},
  \bibinfo{author}{\bibfnamefont{F.}~\bibnamefont{von Oppen}},
  \bibnamefont{and} \bibinfo{author}{\bibfnamefont{S.}~\bibnamefont{Ilani}},
  \bibinfo{journal}{Nature Physics} \textbf{\bibinfo{volume}{10}},
  \bibinfo{pages}{151} (\bibinfo{year}{2014}).

\bibitem[{\citenamefont{Shulaker et~al.}(2013)\citenamefont{Shulaker, Hills,
  Patil, Wei, Chen, Wong, and Mitra}}]{Shulaker:2013}
\bibinfo{author}{\bibfnamefont{M.~M.} \bibnamefont{Shulaker}},
  \bibinfo{author}{\bibfnamefont{G.}~\bibnamefont{Hills}},
  \bibinfo{author}{\bibfnamefont{N.}~\bibnamefont{Patil}},
  \bibinfo{author}{\bibfnamefont{H.}~\bibnamefont{Wei}},
  \bibinfo{author}{\bibfnamefont{H.-Y.} \bibnamefont{Chen}},
  \bibinfo{author}{\bibfnamefont{H.-S.~P.} \bibnamefont{Wong}},
  \bibnamefont{and} \bibinfo{author}{\bibfnamefont{S.}~\bibnamefont{Mitra}},
  \bibinfo{journal}{Nature} \textbf{\bibinfo{volume}{501}},
  \bibinfo{pages}{526} (\bibinfo{year}{2013}).

\bibitem[{\citenamefont{Aurich et~al.}(2010)\citenamefont{Aurich, Baumgartner,
  Freitag, Eichler, Trbovic, and Sch\"onenberger}}]{Aurich:2010}
\bibinfo{author}{\bibfnamefont{H.}~\bibnamefont{Aurich}},
  \bibinfo{author}{\bibfnamefont{A.}~\bibnamefont{Baumgartner}},
  \bibinfo{author}{\bibfnamefont{F.}~\bibnamefont{Freitag}},
  \bibinfo{author}{\bibfnamefont{A.}~\bibnamefont{Eichler}},
  \bibinfo{author}{\bibfnamefont{J.}~\bibnamefont{Trbovic}}, \bibnamefont{and}
  \bibinfo{author}{\bibfnamefont{C.}~\bibnamefont{Sch\"onenberger}},
  \bibinfo{journal}{Applied Physics Letters} \textbf{\bibinfo{volume}{97}},
  \bibinfo{pages}{153116} (\bibinfo{year}{2010}).

\bibitem[{\citenamefont{Sahoo et~al.}(2005)\citenamefont{Sahoo, Kontos, Furer,
  Hoffmann, Gr\"aber, Cottet, and Sch\"onenberger}}]{Sahoo:2005}
\bibinfo{author}{\bibfnamefont{S.}~\bibnamefont{Sahoo}},
  \bibinfo{author}{\bibfnamefont{T.}~\bibnamefont{Kontos}},
  \bibinfo{author}{\bibfnamefont{J.}~\bibnamefont{Furer}},
  \bibinfo{author}{\bibfnamefont{C.}~\bibnamefont{Hoffmann}},
  \bibinfo{author}{\bibfnamefont{M.}~\bibnamefont{Gr\"aber}},
  \bibinfo{author}{\bibfnamefont{A.}~\bibnamefont{Cottet}}, \bibnamefont{and}
  \bibinfo{author}{\bibfnamefont{C.}~\bibnamefont{Sch\"onenberger}},
  \bibinfo{journal}{Nature Physics} \textbf{\bibinfo{volume}{1}},
  \bibinfo{pages}{99} (\bibinfo{year}{2005}).

\bibitem[{\citenamefont{Sch\"onenberger
  et~al.}(1999)\citenamefont{Sch\"onenberger, Bachtold, Strunk, Salvetat, and
  Forr\'{o}}}]{Schonenberger:1999}
\bibinfo{author}{\bibfnamefont{C.}~\bibnamefont{Sch\"onenberger}},
  \bibinfo{author}{\bibfnamefont{A.}~\bibnamefont{Bachtold}},
  \bibinfo{author}{\bibfnamefont{C.}~\bibnamefont{Strunk}},
  \bibinfo{author}{\bibfnamefont{J.-P.} \bibnamefont{Salvetat}},
  \bibnamefont{and}
  \bibinfo{author}{\bibfnamefont{L.}~\bibnamefont{Forr\'{o}}},
  \bibinfo{journal}{Applied Physics A} \textbf{\bibinfo{volume}{69}},
  \bibinfo{pages}{283} (\bibinfo{year}{1999}).

\bibitem[{\citenamefont{Langer et~al.}(1996)\citenamefont{Langer, Bayot,
  Grivei, Issi, Heremans, Olk, Stockman, Van~Haesendonck, and
  Bruynseraede}}]{Langer:1996}
\bibinfo{author}{\bibfnamefont{L.}~\bibnamefont{Langer}},
  \bibinfo{author}{\bibfnamefont{V.}~\bibnamefont{Bayot}},
  \bibinfo{author}{\bibfnamefont{E.}~\bibnamefont{Grivei}},
  \bibinfo{author}{\bibfnamefont{J.-P.} \bibnamefont{Issi}},
  \bibinfo{author}{\bibfnamefont{J.~P.} \bibnamefont{Heremans}},
  \bibinfo{author}{\bibfnamefont{C.~H.} \bibnamefont{Olk}},
  \bibinfo{author}{\bibfnamefont{L.}~\bibnamefont{Stockman}},
  \bibinfo{author}{\bibfnamefont{C.}~\bibnamefont{Van~Haesendonck}},
  \bibnamefont{and}
  \bibinfo{author}{\bibfnamefont{Y.}~\bibnamefont{Bruynseraede}},
  \bibinfo{journal}{Physical Review Letters} \textbf{\bibinfo{volume}{76}},
  \bibinfo{pages}{479} (\bibinfo{year}{1996}).

\bibitem[{\citenamefont{Bockrath et~al.}(1999)\citenamefont{Bockrath, Cobden,
  Lu, Rinzler, Smalley, Balents, and McEuen}}]{Bockrath:1999}
\bibinfo{author}{\bibfnamefont{M.}~\bibnamefont{Bockrath}},
  \bibinfo{author}{\bibfnamefont{D.~H.} \bibnamefont{Cobden}},
  \bibinfo{author}{\bibfnamefont{J.}~\bibnamefont{Lu}},
  \bibinfo{author}{\bibfnamefont{A.~G.} \bibnamefont{Rinzler}},
  \bibinfo{author}{\bibfnamefont{R.~E.} \bibnamefont{Smalley}},
  \bibinfo{author}{\bibfnamefont{L.}~\bibnamefont{Balents}}, \bibnamefont{and}
  \bibinfo{author}{\bibfnamefont{P.~L.} \bibnamefont{McEuen}},
  \bibinfo{journal}{Nature} \textbf{\bibinfo{volume}{397}},
  \bibinfo{pages}{598} (\bibinfo{year}{1999}).

\bibitem[{\citenamefont{Klinovaja and Loss}(2013)}]{Klinovaja:2013}
\bibinfo{author}{\bibfnamefont{J.}~\bibnamefont{Klinovaja}} \bibnamefont{and}
  \bibinfo{author}{\bibfnamefont{D.}~\bibnamefont{Loss}},
  \bibinfo{journal}{Physical Review Letters} \textbf{\bibinfo{volume}{110}},
  \bibinfo{pages}{126402} (\bibinfo{year}{2013}).

\bibitem[{\citenamefont{Miyamoto et~al.}(2001)\citenamefont{Miyamoto, Saito,
  and Tom\'{a}nek}}]{Miyamoto:2001}
\bibinfo{author}{\bibfnamefont{Y.}~\bibnamefont{Miyamoto}},
  \bibinfo{author}{\bibfnamefont{S.}~\bibnamefont{Saito}}, \bibnamefont{and}
  \bibinfo{author}{\bibfnamefont{D.}~\bibnamefont{Tom\'{a}nek}},
  \bibinfo{journal}{Physical Review B} \textbf{\bibinfo{volume}{65}},
  \bibinfo{pages}{041402} (\bibinfo{year}{2001}).

\bibitem[{\citenamefont{van~der Wiel et~al.}(2002)\citenamefont{van~der Wiel,
  De~Franceschi, Elzerman, Fujisawa, Tarucha, and Kouwenhoven}}]{Wiel:2002}
\bibinfo{author}{\bibfnamefont{W.~G.} \bibnamefont{van~der Wiel}},
  \bibinfo{author}{\bibfnamefont{S.}~\bibnamefont{De~Franceschi}},
  \bibinfo{author}{\bibfnamefont{J.~M.} \bibnamefont{Elzerman}},
  \bibinfo{author}{\bibfnamefont{T.}~\bibnamefont{Fujisawa}},
  \bibinfo{author}{\bibfnamefont{S.}~\bibnamefont{Tarucha}}, \bibnamefont{and}
  \bibinfo{author}{\bibfnamefont{L.~P.} \bibnamefont{Kouwenhoven}},
  \bibinfo{journal}{Reviews of Modern Physics} \textbf{\bibinfo{volume}{75}},
  \bibinfo{pages}{1} (\bibinfo{year}{2002}).

\bibitem[{\citenamefont{H\"uttel et~al.}(2006)\citenamefont{H\"uttel, Ludwig,
  Lorenz, Eberl, and Kotthaus}}]{Huttel:2006}
\bibinfo{author}{\bibfnamefont{A.~K.} \bibnamefont{H\"uttel}},
  \bibinfo{author}{\bibfnamefont{S.}~\bibnamefont{Ludwig}},
  \bibinfo{author}{\bibfnamefont{H.}~\bibnamefont{Lorenz}},
  \bibinfo{author}{\bibfnamefont{K.}~\bibnamefont{Eberl}}, \bibnamefont{and}
  \bibinfo{author}{\bibfnamefont{J.~P.} \bibnamefont{Kotthaus}},
  \bibinfo{journal}{Physica E: Low-dimensional Systems and Nanostructures}
  \textbf{\bibinfo{volume}{34}}, \bibinfo{pages}{488} (\bibinfo{year}{2006}).

\bibitem[{\citenamefont{Oosterkamp et~al.}(1998)\citenamefont{Oosterkamp,
  Fujisawa, van~der Wiel, Ishibashi, Hijman, Tarucha, and
  Kouwenhoven}}]{Oosterkamp:1998}
\bibinfo{author}{\bibfnamefont{T.~H.} \bibnamefont{Oosterkamp}},
  \bibinfo{author}{\bibfnamefont{T.}~\bibnamefont{Fujisawa}},
  \bibinfo{author}{\bibfnamefont{W.~G.} \bibnamefont{van~der Wiel}},
  \bibinfo{author}{\bibfnamefont{K.}~\bibnamefont{Ishibashi}},
  \bibinfo{author}{\bibfnamefont{R.~V.} \bibnamefont{Hijman}},
  \bibinfo{author}{\bibfnamefont{S.}~\bibnamefont{Tarucha}}, \bibnamefont{and}
  \bibinfo{author}{\bibfnamefont{L.~P.} \bibnamefont{Kouwenhoven}},
  \bibinfo{journal}{Nature} \textbf{\bibinfo{volume}{395}},
  \bibinfo{pages}{873} (\bibinfo{year}{1998}).

\bibitem[{\citenamefont{Jung et~al.}(2013)\citenamefont{Jung, Schindele, Nau,
  Weiss, Baumgartner, and Sch\"onenberger}}]{Jung:2013}
\bibinfo{author}{\bibfnamefont{M.}~\bibnamefont{Jung}},
  \bibinfo{author}{\bibfnamefont{J.}~\bibnamefont{Schindele}},
  \bibinfo{author}{\bibfnamefont{S.}~\bibnamefont{Nau}},
  \bibinfo{author}{\bibfnamefont{M.}~\bibnamefont{Weiss}},
  \bibinfo{author}{\bibfnamefont{A.}~\bibnamefont{Baumgartner}},
  \bibnamefont{and}
  \bibinfo{author}{\bibfnamefont{C.}~\bibnamefont{Sch\"onenberger}},
  \bibinfo{journal}{Nano Letters} \textbf{\bibinfo{volume}{13}},
  \bibinfo{pages}{4522} (\bibinfo{year}{2013}).

\bibitem[{\citenamefont{Gr\"aber et~al.}(2006)\citenamefont{Gr\"aber, Coish,
  Hoffmann, Weiss, Furer, Oberholzer, Loss, and Sch\"onenberger}}]{Graber:2006}
\bibinfo{author}{\bibfnamefont{M.~R.} \bibnamefont{Gr\"aber}},
  \bibinfo{author}{\bibfnamefont{W.~A.} \bibnamefont{Coish}},
  \bibinfo{author}{\bibfnamefont{C.}~\bibnamefont{Hoffmann}},
  \bibinfo{author}{\bibfnamefont{M.}~\bibnamefont{Weiss}},
  \bibinfo{author}{\bibfnamefont{J.}~\bibnamefont{Furer}},
  \bibinfo{author}{\bibfnamefont{S.}~\bibnamefont{Oberholzer}},
  \bibinfo{author}{\bibfnamefont{D.}~\bibnamefont{Loss}}, \bibnamefont{and}
  \bibinfo{author}{\bibfnamefont{C.}~\bibnamefont{Sch\"onenberger}},
  \bibinfo{journal}{Physical Review B} \textbf{\bibinfo{volume}{74}},
  \bibinfo{pages}{075427} (\bibinfo{year}{2006}).

\bibitem[{\citenamefont{Sapmaz et~al.}(2006)\citenamefont{Sapmaz, Meyer,
  Beliczynski, Jarillo-Herrero, and Kouwenhoven}}]{Sapmaz:2006}
\bibinfo{author}{\bibfnamefont{S.}~\bibnamefont{Sapmaz}},
  \bibinfo{author}{\bibfnamefont{C.}~\bibnamefont{Meyer}},
  \bibinfo{author}{\bibfnamefont{P.}~\bibnamefont{Beliczynski}},
  \bibinfo{author}{\bibfnamefont{P.}~\bibnamefont{Jarillo-Herrero}},
  \bibnamefont{and} \bibinfo{author}{\bibfnamefont{L.~P.}
  \bibnamefont{Kouwenhoven}}, \bibinfo{journal}{Nano Letters}
  \textbf{\bibinfo{volume}{6}}, \bibinfo{pages}{1350} (\bibinfo{year}{2006}).

\bibitem[{\citenamefont{Buitelaar et~al.}(2008)\citenamefont{Buitelaar,
  Fransson, Cantone, Smith, Anderson, Jones, Ardavan, Khlobystov, Watt,
  Porfyrakis et~al.}}]{Buitelaar:2008}
\bibinfo{author}{\bibfnamefont{M.~R.} \bibnamefont{Buitelaar}},
  \bibinfo{author}{\bibfnamefont{J.}~\bibnamefont{Fransson}},
  \bibinfo{author}{\bibfnamefont{A.~L.} \bibnamefont{Cantone}},
  \bibinfo{author}{\bibfnamefont{C.~G.} \bibnamefont{Smith}},
  \bibinfo{author}{\bibfnamefont{D.}~\bibnamefont{Anderson}},
  \bibinfo{author}{\bibfnamefont{G.~A.~C.} \bibnamefont{Jones}},
  \bibinfo{author}{\bibfnamefont{A.}~\bibnamefont{Ardavan}},
  \bibinfo{author}{\bibfnamefont{A.~N.} \bibnamefont{Khlobystov}},
  \bibinfo{author}{\bibfnamefont{A.~A.~R.} \bibnamefont{Watt}},
  \bibinfo{author}{\bibfnamefont{K.}~\bibnamefont{Porfyrakis}},
  \bibnamefont{et~al.}, \bibinfo{journal}{Physical Review B}
  \textbf{\bibinfo{volume}{77}}, \bibinfo{pages}{245439}
  (\bibinfo{year}{2008}).

\bibitem[{\citenamefont{Liu et~al.}(2008)\citenamefont{Liu, Fujisawa, Ono,
  Inokawa, Fujiwara, Takashina, and Hirayama}}]{Liu:2008}
\bibinfo{author}{\bibfnamefont{H.~W.} \bibnamefont{Liu}},
  \bibinfo{author}{\bibfnamefont{T.}~\bibnamefont{Fujisawa}},
  \bibinfo{author}{\bibfnamefont{Y.}~\bibnamefont{Ono}},
  \bibinfo{author}{\bibfnamefont{H.}~\bibnamefont{Inokawa}},
  \bibinfo{author}{\bibfnamefont{A.}~\bibnamefont{Fujiwara}},
  \bibinfo{author}{\bibfnamefont{K.}~\bibnamefont{Takashina}},
  \bibnamefont{and} \bibinfo{author}{\bibfnamefont{Y.}~\bibnamefont{Hirayama}},
  \bibinfo{journal}{Physical Review B} \textbf{\bibinfo{volume}{77}},
  \bibinfo{pages}{073310} (\bibinfo{year}{2008}).

\bibitem[{\citenamefont{Johnson et~al.}(2005)\citenamefont{Johnson, Petta,
  Marcus, Hanson, and Gossard}}]{Johnson:2005}
\bibinfo{author}{\bibfnamefont{A.~C.} \bibnamefont{Johnson}},
  \bibinfo{author}{\bibfnamefont{J.~R.} \bibnamefont{Petta}},
  \bibinfo{author}{\bibfnamefont{C.~M.} \bibnamefont{Marcus}},
  \bibinfo{author}{\bibfnamefont{M.~P.} \bibnamefont{Hanson}},
  \bibnamefont{and} \bibinfo{author}{\bibfnamefont{A.~C.}
  \bibnamefont{Gossard}}, \bibinfo{journal}{Physical Review B}
  \textbf{\bibinfo{volume}{72}}, \bibinfo{pages}{165308}
  (\bibinfo{year}{2005}).

\bibitem[{\citenamefont{Laird et~al.}(2013)\citenamefont{Laird, Pei, and
  Kouwenhoven}}]{Laird:2013}
\bibinfo{author}{\bibfnamefont{E.~A.} \bibnamefont{Laird}},
  \bibinfo{author}{\bibfnamefont{F.}~\bibnamefont{Pei}}, \bibnamefont{and}
  \bibinfo{author}{\bibfnamefont{L.~P.} \bibnamefont{Kouwenhoven}},
  \bibinfo{journal}{Nature Nanotechnology} \textbf{\bibinfo{volume}{8}},
  \bibinfo{pages}{565} (\bibinfo{year}{2013}).

\bibitem[{\citenamefont{Wang et~al.}(2011)\citenamefont{Wang, Guo, Wei, Cao,
  Tu, Xiao, Guo, and Chang}}]{Wang:2011}
\bibinfo{author}{\bibfnamefont{L.-J.} \bibnamefont{Wang}},
  \bibinfo{author}{\bibfnamefont{G.-P.} \bibnamefont{Guo}},
  \bibinfo{author}{\bibfnamefont{D.}~\bibnamefont{Wei}},
  \bibinfo{author}{\bibfnamefont{G.}~\bibnamefont{Cao}},
  \bibinfo{author}{\bibfnamefont{T.}~\bibnamefont{Tu}},
  \bibinfo{author}{\bibfnamefont{M.}~\bibnamefont{Xiao}},
  \bibinfo{author}{\bibfnamefont{G.-C.} \bibnamefont{Guo}}, \bibnamefont{and}
  \bibinfo{author}{\bibfnamefont{A.~M.} \bibnamefont{Chang}},
  \bibinfo{journal}{Applied Physics Letters} \textbf{\bibinfo{volume}{99}},
  \bibinfo{pages}{112117} (\bibinfo{year}{2011}).

\bibitem[{\citenamefont{Datta et~al.}(2011)\citenamefont{Datta, Wang, Tilmaciu,
  Flahaut, Marty, Grifoni, and Wernsdorfer}}]{Datta:2011}
\bibinfo{author}{\bibfnamefont{S.}~\bibnamefont{Datta}},
  \bibinfo{author}{\bibfnamefont{S.}~\bibnamefont{Wang}},
  \bibinfo{author}{\bibfnamefont{C.}~\bibnamefont{Tilmaciu}},
  \bibinfo{author}{\bibfnamefont{E.}~\bibnamefont{Flahaut}},
  \bibinfo{author}{\bibfnamefont{L.}~\bibnamefont{Marty}},
  \bibinfo{author}{\bibfnamefont{M.}~\bibnamefont{Grifoni}}, \bibnamefont{and}
  \bibinfo{author}{\bibfnamefont{W.}~\bibnamefont{Wernsdorfer}},
  \bibinfo{journal}{Physical Review B} \textbf{\bibinfo{volume}{84}},
  \bibinfo{pages}{035408} (\bibinfo{year}{2011}).

\bibitem[{\citenamefont{Moon et~al.}(2007)\citenamefont{Moon, Song, Lee, Kim,
  Kim, Lee, and Choi}}]{Moon:2007}
\bibinfo{author}{\bibfnamefont{S.}~\bibnamefont{Moon}},
  \bibinfo{author}{\bibfnamefont{W.}~\bibnamefont{Song}},
  \bibinfo{author}{\bibfnamefont{J.~S.} \bibnamefont{Lee}},
  \bibinfo{author}{\bibfnamefont{N.}~\bibnamefont{Kim}},
  \bibinfo{author}{\bibfnamefont{J.}~\bibnamefont{Kim}},
  \bibinfo{author}{\bibfnamefont{S.-G.} \bibnamefont{Lee}}, \bibnamefont{and}
  \bibinfo{author}{\bibfnamefont{M.-S.} \bibnamefont{Choi}},
  \bibinfo{journal}{Physical Review Letters} \textbf{\bibinfo{volume}{99}},
  \bibinfo{pages}{176804} (\bibinfo{year}{2007}).

\bibitem[{\citenamefont{Go\ss et~al.}(2013)\citenamefont{Go\ss, Leijnse,
  Smerat, Wegewijs, Schneider, and Meyer}}]{Gos:2013}
\bibinfo{author}{\bibfnamefont{K.}~\bibnamefont{Go\ss}},
  \bibinfo{author}{\bibfnamefont{M.}~\bibnamefont{Leijnse}},
  \bibinfo{author}{\bibfnamefont{S.}~\bibnamefont{Smerat}},
  \bibinfo{author}{\bibfnamefont{M.~R.} \bibnamefont{Wegewijs}},
  \bibinfo{author}{\bibfnamefont{C.~M.} \bibnamefont{Schneider}},
  \bibnamefont{and} \bibinfo{author}{\bibfnamefont{C.}~\bibnamefont{Meyer}},
  \bibinfo{journal}{Physical Review B} \textbf{\bibinfo{volume}{87}},
  \bibinfo{pages}{035424} (\bibinfo{year}{2013}).

\bibitem[{\citenamefont{Kang et~al.}(2015)\citenamefont{Kang, Kim, and
  Kwon}}]{Kang:2015}
\bibinfo{author}{\bibfnamefont{S.-H.} \bibnamefont{Kang}},
  \bibinfo{author}{\bibfnamefont{G.}~\bibnamefont{Kim}}, \bibnamefont{and}
  \bibinfo{author}{\bibfnamefont{Y.-K.} \bibnamefont{Kwon}},
  \bibinfo{journal}{Physical Chemistry Chemical Physics}
  \textbf{\bibinfo{volume}{17}}, \bibinfo{pages}{5072} (\bibinfo{year}{2015}).

\bibitem[{\citenamefont{Samm et~al.}(2014)\citenamefont{Samm, Gramich,
  Baumgartner, Weiss, and Sch\"onenberger}}]{Samm:2014}
\bibinfo{author}{\bibfnamefont{J.}~\bibnamefont{Samm}},
  \bibinfo{author}{\bibfnamefont{J.}~\bibnamefont{Gramich}},
  \bibinfo{author}{\bibfnamefont{A.}~\bibnamefont{Baumgartner}},
  \bibinfo{author}{\bibfnamefont{M.}~\bibnamefont{Weiss}}, \bibnamefont{and}
  \bibinfo{author}{\bibfnamefont{C.}~\bibnamefont{Sch\"onenberger}},
  \bibinfo{journal}{Journal of Applied Physics} \textbf{\bibinfo{volume}{115}},
  \bibinfo{pages}{174309} (\bibinfo{year}{2014}).

\bibitem[{\citenamefont{Hasler et~al.}(2015)\citenamefont{Hasler, Jung, Ranjan,
  Puebla-Hellmann, Wallraff, and Sch\"onenberger}}]{Hasler:2015}
\bibinfo{author}{\bibfnamefont{T.}~\bibnamefont{Hasler}},
  \bibinfo{author}{\bibfnamefont{M.}~\bibnamefont{Jung}},
  \bibinfo{author}{\bibfnamefont{V.}~\bibnamefont{Ranjan}},
  \bibinfo{author}{\bibfnamefont{G.}~\bibnamefont{Puebla-Hellmann}},
  \bibinfo{author}{\bibfnamefont{A.}~\bibnamefont{Wallraff}}, \bibnamefont{and}
  \bibinfo{author}{\bibfnamefont{C.}~\bibnamefont{Sch\"onenberger}},
  \bibinfo{journal}{Physical Review Applied} \textbf{\bibinfo{volume}{4}},
  \bibinfo{pages}{054002} (\bibinfo{year}{2015}).

\bibitem[{\citenamefont{Viennot et~al.}(2014)\citenamefont{Viennot, Palomo, and
  Kontos}}]{Viennot:2014}
\bibinfo{author}{\bibfnamefont{J.~J.} \bibnamefont{Viennot}},
  \bibinfo{author}{\bibfnamefont{J.}~\bibnamefont{Palomo}}, \bibnamefont{and}
  \bibinfo{author}{\bibfnamefont{T.}~\bibnamefont{Kontos}},
  \bibinfo{journal}{Applied Physics Letters} \textbf{\bibinfo{volume}{104}},
  \bibinfo{pages}{113108} (\bibinfo{year}{2014}).

\bibitem[{\citenamefont{Gramich
  et~al.}(2015{\natexlab{b}})\citenamefont{Gramich, Baumgartner, Muoth,
  Hierold, and Sch\"onenberger}}]{Gramich:2015-2}
\bibinfo{author}{\bibfnamefont{J.}~\bibnamefont{Gramich}},
  \bibinfo{author}{\bibfnamefont{A.}~\bibnamefont{Baumgartner}},
  \bibinfo{author}{\bibfnamefont{M.}~\bibnamefont{Muoth}},
  \bibinfo{author}{\bibfnamefont{C.}~\bibnamefont{Hierold}}, \bibnamefont{and}
  \bibinfo{author}{\bibfnamefont{C.}~\bibnamefont{Sch\"onenberger}},
  \bibinfo{journal}{physica status solidi (b)} \textbf{\bibinfo{volume}{252}},
  \bibinfo{pages}{2496} (\bibinfo{year}{2015}{\natexlab{b}}).

\bibitem[{\citenamefont{Huang et~al.}(2015)\citenamefont{Huang, Pan, Tran,
  Cheng, Watanabe, Taniguchi, Lau, and Bockrath}}]{Huang:2015}
\bibinfo{author}{\bibfnamefont{J.-W.} \bibnamefont{Huang}},
  \bibinfo{author}{\bibfnamefont{C.}~\bibnamefont{Pan}},
  \bibinfo{author}{\bibfnamefont{S.}~\bibnamefont{Tran}},
  \bibinfo{author}{\bibfnamefont{B.}~\bibnamefont{Cheng}},
  \bibinfo{author}{\bibfnamefont{K.}~\bibnamefont{Watanabe}},
  \bibinfo{author}{\bibfnamefont{T.}~\bibnamefont{Taniguchi}},
  \bibinfo{author}{\bibfnamefont{C.~N.} \bibnamefont{Lau}}, \bibnamefont{and}
  \bibinfo{author}{\bibfnamefont{M.}~\bibnamefont{Bockrath}},
  \bibinfo{journal}{Nano Letters} \textbf{\bibinfo{volume}{15}},
  \bibinfo{pages}{6836} (\bibinfo{year}{2015}).

\bibitem[{\citenamefont{Baumgartner et~al.}(2014)\citenamefont{Baumgartner,
  Abulizi, Watanabe, Taniguchi, Gramich, and
  Sch\"onenberger}}]{Baumgartner:2014}
\bibinfo{author}{\bibfnamefont{A.}~\bibnamefont{Baumgartner}},
  \bibinfo{author}{\bibfnamefont{G.}~\bibnamefont{Abulizi}},
  \bibinfo{author}{\bibfnamefont{K.}~\bibnamefont{Watanabe}},
  \bibinfo{author}{\bibfnamefont{T.}~\bibnamefont{Taniguchi}},
  \bibinfo{author}{\bibfnamefont{J.}~\bibnamefont{Gramich}}, \bibnamefont{and}
  \bibinfo{author}{\bibfnamefont{C.}~\bibnamefont{Sch\"onenberger}},
  \bibinfo{journal}{Applied Physics Letters} \textbf{\bibinfo{volume}{105}},
  \bibinfo{pages}{023111} (\bibinfo{year}{2014}).

\bibitem[{\citenamefont{Li et~al.}(2009)\citenamefont{Li, Cai, An, Kim, Nah,
  Yang, Piner, Velamakanni, Jung, Tutuc et~al.}}]{Li:2009}
\bibinfo{author}{\bibfnamefont{X.}~\bibnamefont{Li}},
  \bibinfo{author}{\bibfnamefont{W.}~\bibnamefont{Cai}},
  \bibinfo{author}{\bibfnamefont{J.}~\bibnamefont{An}},
  \bibinfo{author}{\bibfnamefont{S.}~\bibnamefont{Kim}},
  \bibinfo{author}{\bibfnamefont{J.}~\bibnamefont{Nah}},
  \bibinfo{author}{\bibfnamefont{D.}~\bibnamefont{Yang}},
  \bibinfo{author}{\bibfnamefont{R.}~\bibnamefont{Piner}},
  \bibinfo{author}{\bibfnamefont{A.}~\bibnamefont{Velamakanni}},
  \bibinfo{author}{\bibfnamefont{I.}~\bibnamefont{Jung}},
  \bibinfo{author}{\bibfnamefont{E.}~\bibnamefont{Tutuc}},
  \bibnamefont{et~al.}, \bibinfo{journal}{Science}
  \textbf{\bibinfo{volume}{324}}, \bibinfo{pages}{1312} (\bibinfo{year}{2009}).

\bibitem[{\citenamefont{Longchamp et~al.}(2013)\citenamefont{Longchamp, Escher,
  and Fink}}]{Longchamp:2013}
\bibinfo{author}{\bibfnamefont{J.-N.} \bibnamefont{Longchamp}},
  \bibinfo{author}{\bibfnamefont{C.}~\bibnamefont{Escher}}, \bibnamefont{and}
  \bibinfo{author}{\bibfnamefont{H.-W.} \bibnamefont{Fink}},
  \bibinfo{journal}{Journal of Vacuum Science \& Technology B}
  \textbf{\bibinfo{volume}{31}}, \bibinfo{pages}{020605}
  (\bibinfo{year}{2013}).

\bibitem[{\citenamefont{Kouwenhoven et~al.}(2001)\citenamefont{Kouwenhoven,
  Austing, and Tarucha}}]{Kouwenhoven:2001}
\bibinfo{author}{\bibfnamefont{L.~P.} \bibnamefont{Kouwenhoven}},
  \bibinfo{author}{\bibfnamefont{D.~G.} \bibnamefont{Austing}},
  \bibnamefont{and} \bibinfo{author}{\bibfnamefont{S.}~\bibnamefont{Tarucha}},
  \bibinfo{journal}{Reports on Progress in Physics}
  \textbf{\bibinfo{volume}{64}}, \bibinfo{pages}{701} (\bibinfo{year}{2001}).

\bibitem[{\citenamefont{Lemay et~al.}(2001)\citenamefont{Lemay, Janssen,
  van~den Hout, Mooij, Bronikowski, Willis, Smalley, Kouwenhoven, and
  Dekker}}]{Lemay:2001}
\bibinfo{author}{\bibfnamefont{S.~G.} \bibnamefont{Lemay}},
  \bibinfo{author}{\bibfnamefont{J.~W.} \bibnamefont{Janssen}},
  \bibinfo{author}{\bibfnamefont{M.}~\bibnamefont{van~den Hout}},
  \bibinfo{author}{\bibfnamefont{M.}~\bibnamefont{Mooij}},
  \bibinfo{author}{\bibfnamefont{M.~J.} \bibnamefont{Bronikowski}},
  \bibinfo{author}{\bibfnamefont{P.~A.} \bibnamefont{Willis}},
  \bibinfo{author}{\bibfnamefont{R.~E.} \bibnamefont{Smalley}},
  \bibinfo{author}{\bibfnamefont{L.~P.} \bibnamefont{Kouwenhoven}},
  \bibnamefont{and} \bibinfo{author}{\bibfnamefont{C.}~\bibnamefont{Dekker}},
  \bibinfo{journal}{Nature} \textbf{\bibinfo{volume}{412}},
  \bibinfo{pages}{617} (\bibinfo{year}{2001}).

\bibitem[{\citenamefont{Cao et~al.}(2005)\citenamefont{Cao, Wang, and
  Dai}}]{Cao:2005-2}
\bibinfo{author}{\bibfnamefont{J.}~\bibnamefont{Cao}},
  \bibinfo{author}{\bibfnamefont{Q.}~\bibnamefont{Wang}}, \bibnamefont{and}
  \bibinfo{author}{\bibfnamefont{H.}~\bibnamefont{Dai}},
  \bibinfo{journal}{Nature Materials} \textbf{\bibinfo{volume}{4}},
  \bibinfo{pages}{745} (\bibinfo{year}{2005}).

\end{thebibliography}

\end{document}